\documentclass[amsmath]{revtex4}
\usepackage{times}
\usepackage{graphicx}
\usepackage{bm}
\usepackage{wasysym}
\usepackage{amssymb}

\begin{document}


\title[Gaped graphene bilayer]{Gaped graphene bilayer:
disorder and magnetic field effects}

\author{Eduardo V. Castro$^1$, N.~M.~R.~Peres$^2$, 
  J.~M.~B.~Lopes~dos~Santos$^1$}

\affiliation{$^1$CFP and Departamento de F\'{\i}sica, 
  Faculdade de Ci\^encias Universidade do Porto, P-4169-007 Porto, Portugal}

\affiliation{$^2$Center of Physics and Departamento de
  F\'{\i}sica, Universidade do Minho, P-4710-057, Braga, Portugal}

\begin{abstract}
Double layer graphene is a gapless semiconductor which develops a finite
gap when the layers are placed at different electrostatic potentials.
We study, within the tight-biding approximation, the electronic properties
of the gaped graphene bilayer in the presence of disorder, perpendicular 
magnetic field, and transverse electric field. We show that the gap is
rather stable in the presence of diagonal disorder. We compute the
cyclotron effective mass in the semi-classical approximation, valid at
low magnetic fields. Landau level formation is clearly seen in zigzag and 
armchair ribbons of the gaped bilayer at intermediate magnetic fields. 
\end{abstract}

\maketitle

\section{Introduction}
\label{sec:intro}
Graphene, the two-dimensional allotropic form of carbon, has recently 
deserved considerable attention. The realization that graphene could be 
obtained and studied experimentally revealed a plethora of unusual 
properties that may be useful in the design of new electronic 
devices [1--3]. The peculiar band structure of
graphene, where at the corners of the Brillouin zone (BZ) the excitations
are massless Dirac fermions, is responsible for many of the unconventional 
properties \cite{nunoPRB}.

The possibility of creating stacks of graphene layers with the accuracy of
a single atomic layer, providing an extra dimension to be explored, is 
another advantage of graphene for electronic applications.
In particular, bilayer graphene has shown to have unusual electronic 
properties, though unexpectedly dissimilar to those exhibited by its
single layer parent. The new type of integer quantum Hall effect
observed in bilayer graphene \cite{NMcCM+06, MF06}, which is induced by
chiral parabolic bands, is an example of its uniqueness. 
Recent advances on the experimental side made it possible to produce bilayers
where the two layers are effectively at different electrostatic potentials
\cite{privateGeim, arpes} -- introducing the concept of 
{\it biased bilayer}. This asymmetry between layers opens a gap 
between valence and conduction bands, which, in the unbiased case (both layers
at the same potential), touch in a parabolic way at zero energy. The gap is a
function of the potential between layers, and may be controlled by changing
the electric field across the bilayer. This switching functionality opens the
door for potential applications of bilayer graphene on atomic-scale electronic 
devices \cite{johan}.

In this paper we study the biased bilayer system starting from the simplest
tight-binding model. By computing the density of states (DOS) we analyze how 
the gap structure evolves when diagonal disorder is present. 
In the presence of a perpendicular magnetic
field we obtain the cyclotron effective mass (semi-classical approximation), 
and study the Landau level formation in 
zigzag and armchair ribbons of bilayer graphene.
The paper is organized as follows: in Sec.~\ref{sec:model} we present the
tight-binding Hamiltonian we use and discuss basic aspects of the electronic
structure; the effect of disorder in the gap structure is discussed
in Sec.~\ref{sec:disorder}; in Sec.~\ref{sec:magnetic} we analyze the effects
of a perpendicular magnetic field; and Sec.~\ref{sec:conclusions} contains our conclusions.

\section{Model and basic electronic structure}
\label{sec:model}
In single layer graphene the carbon atoms form a honeycomb lattice which has
two atoms per unit cell. These two atoms belong to different triangular 
lattices labeled~A and~B. The bilayer is made of two layers, which we label~1
and~2, arranged in the Bernal stacking (A1-B2). The unit cell has then four
atoms, one per layer per sublattice. Each carbon atom contributes with a $\pi$
electron, and thus the system is naturally at half-filling.

In the tight-binding approximation the relevant energy scales are the
in-plane hopping energy, $t\approx 2.7\,\textrm{eV}$, and the interlayer 
hopping energy, $t_{\perp} \approx 0.35\,\textrm{eV}$.
 When a perpendicular magnetic field 
$\mathbf{B} = B\,\textrm{\^e}_z$ is applied to the system
$t_{\perp}$ is unaffected but $t$ acquires a phase \cite{Peierls33}
such that, $t \rightarrow te^{i \frac{e}{\hbar} 
\int_{\mathbf{R}}^{\mathbf{R}+\bm\delta}
\mathbf{A}\cdot\textrm{d}\mathbf{r}}$, where $e$ is the electron charge,
$\bm \delta$ is a vector connecting nearest-neighbor sites, and $\mathbf{A}$ is the vector potential.
The tight-binging Hamiltonian describing non-interacting $\pi$~electrons in 
bilayer graphene then reads:
\begin{equation} \label{eq:Hbilayer}
\begin{split}
&H =  - t\sum_{m,n}\big[
e^{i\pi \frac{\phi}{\phi_0} n}
a_{1,(m,n)}^{\dagger}b_{1,(m,n)}
+e^{- i\pi \frac{\phi}{\phi_0} n}
a_{1,(m,n)}^{\dagger}b_{1,(m-1,n)}
+a_{1,(m,n)}^{\dagger}b_{1,(m,n-1)}
+\textrm{h.c.}\big] \\
& - t\sum_{m,n}\big[
e^{- i\pi \frac{\phi}{\phi_0} (n - \frac{1}{3})}
b_{2,(m,n)}^{\dagger}a_{2,(m,n)}
+e^{i\pi \frac{\phi}{\phi_0} (n - \frac{1}{3})}
 b_{2,(m,n)}^{\dagger}a_{2,(m+1,n)}
+ b_{2,(m,n)}^{\dagger}a_{2,(m,n+1)}
+\textrm{h.c.}\big] \\
& - t_{\perp}\sum_{m,n}\big[
a_{1,(m,n)}^{\dagger}b_{2,(m,n)}+\textrm{h.c.}\big]
+ \frac{V}{2}\sum_{m,n}\big[
n_{\textrm{A}1,(m,n)}+n_{\textrm{B}1,(m,n)}
-n_{\textrm{A}2,(m,n)}-n_{\textrm{B}2,(m,n)}\big]\,,
\end{split}
\end{equation}
where $a_{i,(m,n)}^{\dagger}$ 
\big[$a_{i,(m,n)}$\big] and
$b_{i,(m,n)}^{\dagger}$ \big[$b_{i,(m,n)}$\big]
creates (annihilates) an electron on atom A$i$ and B$i$ of layer~$i$ at cell
 $(m,n)$, respectively. Spin indices have been neglected for simplicity. 
The Peierls phase in 
Eq.~\eqref{eq:Hbilayer} is written assuming a Landau gauge,
$\mathbf{A} = (-y, 0, 0)B$, where $\phi/ \phi_0=BA_{\hexagon}/ \phi_0$ is
the magnetic flux through a plaquette in units of the flux 
quantum $\phi_0=h/e$ ($A_{\hexagon}$ being the area of the graphene 
unit cell). 
The last term in Eq.~\eqref{eq:Hbilayer} stands for the electrostatic
potential difference, $V$, between the two layers, 
with $n_{Ai,(m,n)}$ and $n_{Bi,(m,n)}$ number operators.

\begin{figure}[t]
\begin{center}
\includegraphics[width=0.6\columnwidth]{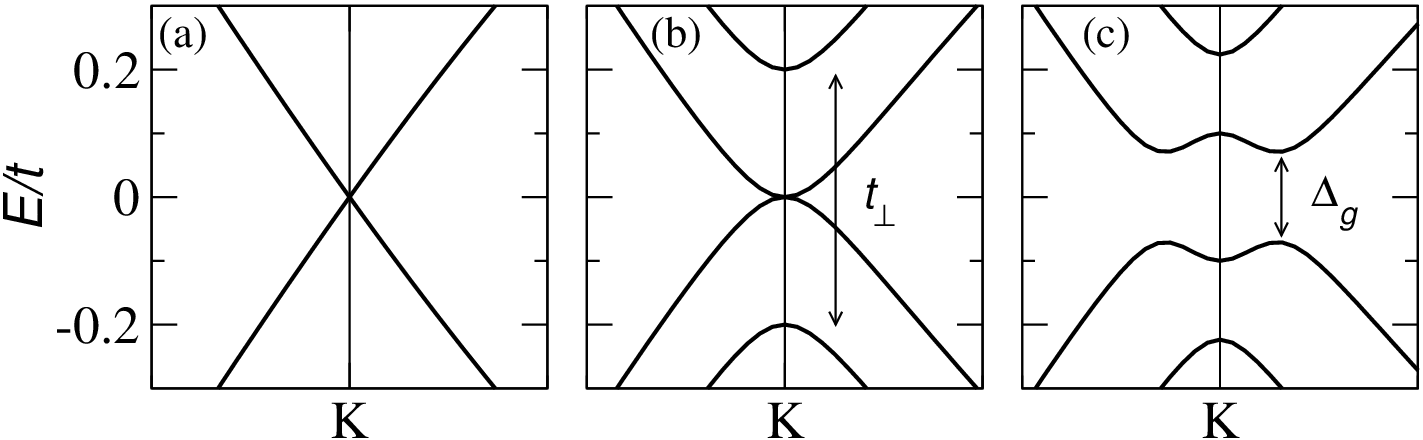}
\end{center}
\caption{\label{cap:fig1} Band structure of
single layer graphene~(a), unbiased bilayer graphene~(b), and
biased bilayer~(c) with $V=2t_{\perp}$,  along two directions in the BZ 
near the Dirac (K) points. The inter-layer hopping was set to
$t_{\perp}/t=0.2$.}
\end{figure}

The first (or second) line of Eq.~\eqref{eq:Hbilayer} describes $\pi$~electrons
on single layer graphene. The resultant dispersion relation is given by,
\begin{equation} \label{eq:Ek1L}
\epsilon_{\mathbf{k}}^2/t^2 = 3 + 2\cos(ak_{x}) + 
4\cos\left(ak_{x}/2\right)\cos\left(ak_{y}\sqrt{3}/2\right)\,,
\end{equation}
where $a$ is the in-plane C-C distance for sites of the same sublattice.
The vertex of the valence and conduction bands touch at the corners of 
the BZ (Dirac points), as shown in Fig.~\ref{cap:fig1}~(a). A Dirac linear
dispersion follows at low energies, $\epsilon(q) = \pm v_{\textrm{F}} \hbar q$,
where $v_{\textrm{F}} = ta\hbar^{-1}\sqrt{3}/2$ is the Fermi velocity which
substitutes the speed of light, and $\mathbf{q}$ is the wave vector relatively
to the Dirac points.
Taking into account the full Hamiltonian given in Eq.~\eqref{eq:Hbilayer}
we obtain four bands with dispersion energies given by,
\begin{equation} \label{eq:Ek2L}
E_{\mathbf{k}}^{\pm\pm}(V) \!\!=\!\!
\pm \sqrt{\!\epsilon_{\mathbf{k}}^{2}  \!+\! t_{\perp}^{2}/2 \!+\!
V^{2}/4 \!\pm\! \sqrt{\!t_{\perp}^{4}/4
+(t_{\perp}^{2}\!+\!V^{2}\!)\epsilon_{\mathbf{k}}^{2}}}\,,
\end{equation}
with $\epsilon_{\mathbf{k}}$ as in Eq.~\eqref{eq:Ek1L}. For the unbiased
case Eq.~\eqref{eq:Ek2L} reduces to 
$E_{\mathbf{k}}^{\pm \pm} = \pm (\epsilon_{\mathbf{k}}^{2}
+ t_{\perp}^2/4)^{1/2} \pm t_{\perp}/2$, from which it becomes clear
that the spectrum is as in Fig.~\ref{cap:fig1}~(b): two low energy 
gapless bands touching in a parabolic way at the Dirac points, and
two high energy bands with a gap of $t_\perp$. As the undoped bilayer
is at half-filling, the Fermi energy ($E_\textrm{F}$) occurs at exactly
at the points the bands $E_{\mathbf{k}}^{\pm -}$ touch. Thus, low energy
quasi-particles are massive Dirac fermions with effective mass given by
$m^{*}=2t_{\perp}\hbar^2/3a^2t^2$ \cite{MF06,NNGP06}.
The biased bilayer, however, shows a true gap between valence and conduction
bands, as shown in Fig.~\ref{cap:fig1}~(c), and the low energy bands have a
{}``Mexican hat'' like dispersion. For $V \ll t$, which is the case for the
biased bilayer, the gap behaves as $\Delta_{g}=
[t_{\perp}^{2}V^2/(t_{\perp}^{2}+V^2)]^{1/2}$, and is therefore fully 
controlled by the bias $V$.

\section{\bf Effect of diagonal disorder in the gaped structure}
\label{sec:disorder}

Recent experiments using angle-resolved photoemission spectroscopy (ARPES)
have clearly shown that the spectrum of the biased bilayer is well described 
by Eq.~\eqref{eq:Ek2L} \cite{arpes}, with the measured energy-momentum 
dispersion in good agreement with Fig.~\ref{cap:fig1}~(c). However, the 
experimental data also shows the presence of a finite spectral weight inside
the gap, which may be attributed to disorder. The sources of disorder in the
biased bilayer are not yet understood. Inhomogeneities in the substrate,
below the bottom layer, as well as in the dopant coverages, above the top
layer, are plausible sources. Here we consider the biased 
bilayer in the presence of diagonal disorder, adding the term
$H_\textrm{dis} = \sum_{m,n,i=1,2} \varepsilon_{i,(m,n)}
[n_{\textrm{A}i,(m,n)} + n_{\textrm{B}i,(m,n)}]$ to Eq.~\eqref{eq:Hbilayer}.
We use the box distribution function to generate the random on-site energies,
$\varepsilon \in [-W/2, W/2]$, and vary its width $W$ in order to induce 
spectral weight inside the unperturbed gap. 

To study how diagonal disorder affects the gap, and in particular how $W$
compares with the other energy scales of the system, we have computed the 
DOS with the recursion method \cite{Hayd80}. We define zero temperature
retarded Green's functions in the standard way,
\begin{equation} \label{eq:GRt}
G_{\mathbf{r}\mathbf{r}'}^{aa,i}(t) = 
-i\left\langle 0\right|\{ a_{i,\mathbf{r}}(t),
 a_{i,\mathbf{r}'}^{\dagger}(0)\}\left|0\right\rangle \Theta(t),
\hspace{0.5cm}
G_{\mathbf{r}\mathbf{r}'}^{bb,i}(t) = 
-i\left\langle 0\right|\{ b_{i,\mathbf{r}}(t),
 b_{i,\mathbf{r}'}^{\dagger}(0)\}\left|0\right\rangle \Theta(t),
\end{equation}
where $\mathbf{r} = (m,n)$ specifies the cell.
The recursion method gives an approximate value for the retarded Green's
function in the thermodynamic limit,
and therefore for the disordered DOS,
by simulating large lattices. The disorder averaged DOS is then easily
obtained by averaging over disorder realizations.

\begin{figure}[t]
\begin{center}
\includegraphics[width=0.75\columnwidth]{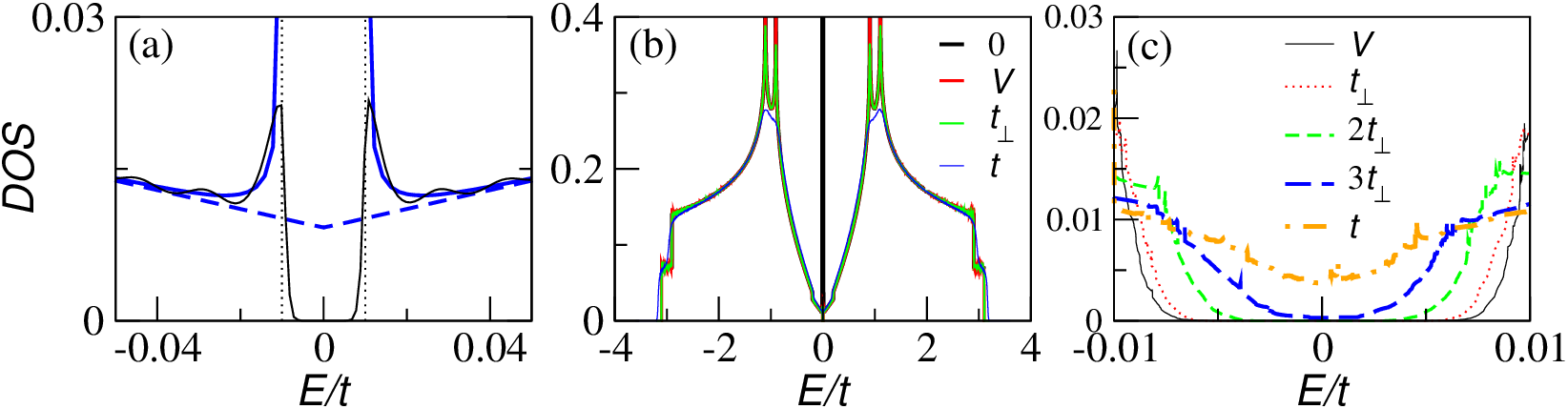}
\end{center}
\caption{\label{cap:fig2}(Color online) (a)~-~Non-disordered DOS: 
exact (thick lines)
and recursion method result (normal line). (b)~-~DOS for several
values of disorder $W$. (c)~-~The same as in~(b) closer to the gap region.}
\end{figure}
Figure~\ref{cap:fig2} resumes the results we have obtained for the effect
of diagonal disorder in the DOS, with emphasis in the gap behavior.
The simulated lattices had $2d^2$ sites, with $d = 1280$, and we averaged
over 100 disorder realizations. The inter-layer
hopping was set to $t_\perp / t = 0.2$. The bias was fixed to $V/t = 0.02$, 
as experimentally it is the smallest energy scale, $V \ll t_\perp$,
and thus the gap is essentially the bias, $\Delta_g \approx V$.
In order to check the performance of the recursion method in describing
such a tiny region of the spectrum ($\Delta_g \ll t_\perp \ll t$), we
show in Fig.~\ref{cap:fig2}~(a) the exact non-disordered DOS \cite{next}
along with the recursion method result. The non-disordered DOS of the
unbiased bilayer (dashed line) falls linearly when $|E| \rightarrow 0$, 
being finite at zero energy \cite{NNGP06,next}. 
When the layers are made inequivalent by the 
bias a gap opens and 1D-like divergences associated with the 
{}``Mexican hat'' like dispersion region show up \cite{GNP06}.
The recursion method reproduces this behavior fairly well, 
in spite of a somewhat
underestimated gap (doted vertical lines signal the exact the gap limits).
In Fig.~\ref{cap:fig2}~(b) we show the DOS for several disorder values.
The disorder parameter $W$ equals the relevant energy scales of
the problem: bias, $V$; inter-plane hopping, $t_\perp$; and in-plane hopping
$t$. It is readily seen that the disordered DOS starts to depart from the 
non-disordered one only for $W \sim t$. Figure~\ref{cap:fig2}~(c) is our
main result of this section, and shows the disordered DOS in the gap 
region. Diagonal disorder closes the gap when its width is of the order of
the largest energy scale in the system, $W \sim t$.

\section{\bf Magnetic field effects}
\label{sec:magnetic}
We have computed the cyclotron effective mass $m_{\textrm{c}}^{*}(n)$, whose
semi-classical expression is 
given by
$m_{\textrm{c}}^{*}(n) = 
(\hbar^{2}/2\pi) \partial A(E)/\partial E |_{E=E_{\textrm{F}}(n)},$
%
where $A(E)$ is the $k$-space area enclosed by the orbit of energy
$E$ and $n$ is the carrier density at the Fermi energy $E_{\textrm{F}}$.
Its dependence on the carrier density $n$ may be found through 
measurements of Shubnikov de Haas oscillations.
Here we consider only the experimentally relevant case $V\ll t_{\perp} \ll t$,
and we further assume that the inequality $E_{\textrm{F}} \ll t$ holds.
In that case the Dirac linear dispersion can be used for 
$\epsilon_\mathbf{k}$ in Eq.~\eqref{eq:Ek2L}, and analytical expressions
for the cyclotron effective mass may be derived. When $E_{\textrm{F}}$ is
varied we have to distinguish three different cases (see Fig.~\ref{cap:fig2}):
for $|E_{\textrm{F}}|$ in the {}``Mexican hat'' region, 
$|E_{\textrm{F}}| < V/2$, there are two types of quasi-particles with Fermi 
wave vectors $q_{\textrm{F}}^{+}$ and $q_{\textrm{F}}^{-}$ (relatively 
to Dirac points) and opposite cyclotron effective masses; 
when $|E_{\textrm{F}}|$ is between the top of the {}``Mexican hat'' and the
bottom of the high energy band, 
$V/2 < E_{\textrm{F}} < (t_\perp^2 + V^2/4)^{1/2}$, only $q_\textrm{F}^+$
quasi-particles exist; $q_\textrm{F}^-$ quasi-particles show up again for 
$E_{\textrm{F}} > (t_\perp^2 + V^2/4)^{1/2}$ with positive cyclotron effective 
mass. The Fermi wave vectors are functions of the Fermi energy,
$taq_{\textrm{F}}^{\pm} = (2/\sqrt{3})\{E_{\textrm{F}}^{2}+V^{2}/4 \pm
[E_{\textrm{F}}^{2}(V^{2}+t_{\perp}^{2}) - 
t_{\perp}^{2}V^{2}/4]^{1/2}\}^{1/2}$,
and the respective cyclotron effective masses are given by,
\begin{equation}
\label{eq:mc}
m_\textrm{c}^{\pm} = 
\frac{\hbar^{2}}{a^{2}t^2} \frac{2}{3} \left[2E_{\textrm{F}} \pm
\frac{E_{\textrm{F}}(V^{2}+t_{\perp}^{2})}
     {\sqrt{E_{\textrm{F}}^{2}(V^{2} + t_{\perp}^{2}) - t_{\perp}^{2}V^{2}/4}}
     \right],\\
\end{equation}
where the Fermi energy depends on the density as,
\begin{equation}
\label{eq:Ef}
E_{\textrm{F}} = 
\begin{cases}
\sqrt{\frac{(\frac{3\pi}{8}\tilde{n})^{2} + \frac{t_{\perp}^{2}V^{2}}{4}}
  {V^{2}+t_{\perp}^{2}}} & \tilde{n} < \frac{4}{3\pi}V^{2}\\
\sqrt{\frac{3\pi}{4} \tilde{n} + \frac{t_{\perp}^{2}}{2} + \frac{V^{2}}{4} - 
  \sqrt{\frac{t_{\perp}^{4}}{4} + 
    (t_{\perp}^{2} + V^{2})\frac{3\pi}{4} \tilde{n}}} & 
\frac{4}{3\pi }V^{2} < \tilde{n} < 
\frac{4}{3\pi }(2t_{\perp}^{2}+V^{2})\\
\sqrt{\frac{3}{8} \pi \tilde{n} - \frac{V^{2}}{4}} & 
\tilde{n} > \frac{4}{3\pi }(2t_{\perp}^{2}+V^{2})
\end{cases}\,,
\end{equation}
with $\tilde{n} = a^2 t^2 n$.
\begin{figure}[t]
  \begin{minipage}[t]{.425\textwidth} 
\includegraphics[width=\textwidth]{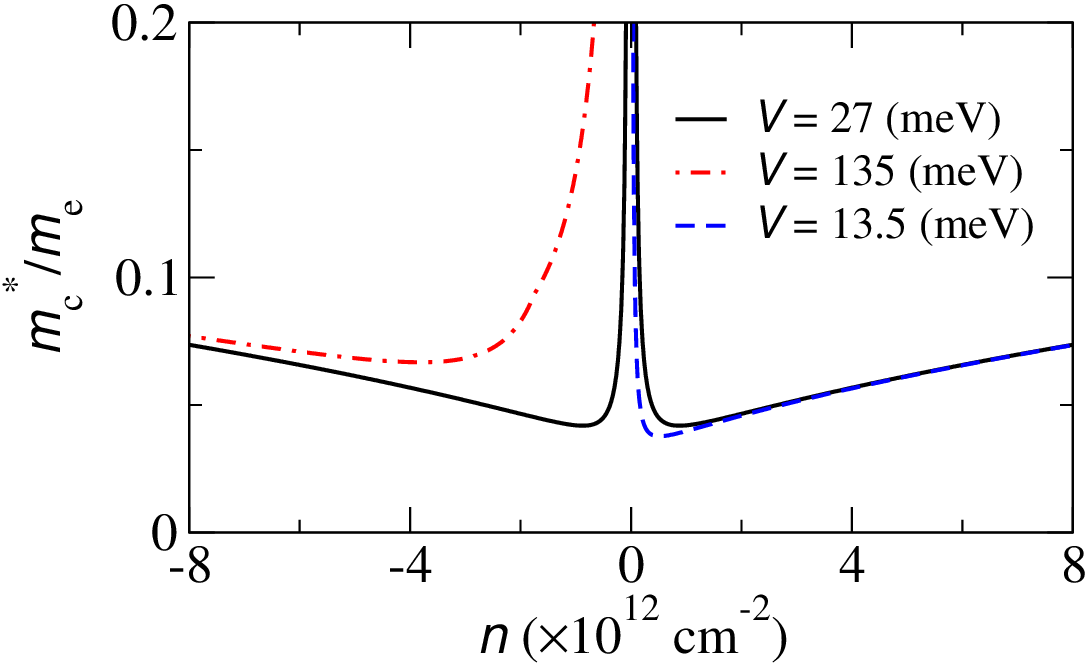} 
\caption{\label{cap:fig3}(Color online)
 Cyclotron effective mass $m_\textrm{c}^+$ 
[given in Eq.~\eqref{eq:mc}] in units of the bare electron mass, 
$m_\textrm{e}$, as a function of the carrier density for three different bias
values. We set $t_{\perp}/t=0.1$}
\end{minipage} 
\hfil 
\begin{minipage}[t]{.475\textwidth} 
\includegraphics[width=\textwidth]
		{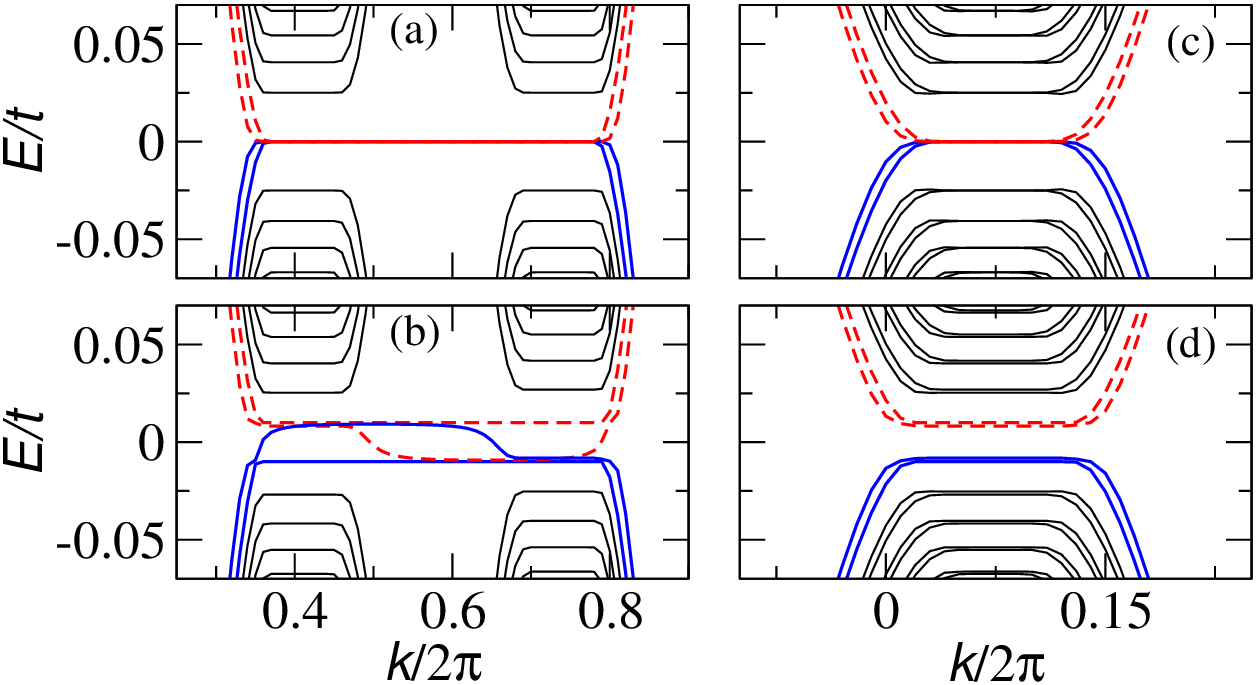} 
\caption{\label{cap:fig4}(Color online)
 Energy spectrum for a ribbon of bilayer
graphene with zigzag~(a-b) and armchair~(c-d) edges and width $N=400$ 
unit cells:
(a-c)~--~ $V=0$;  (b-d)~--~ $V=t_{\perp}/10$.
We set \textbf{$B=30\,$}T and $t_{\perp}/t=0.2$.} 
\end{minipage} 
\end{figure}
Figure~\ref{cap:fig3} shows the result for $m_\textrm{c}^+$ [Eq.~\eqref{eq:mc}]
as a function of the carrier density; $n < 0$ for holes and $n > 0$ for
electrons. As can be seen for 
$V/t = 0.01 \approx 27$~(meV)~--~full line~--~there is a $1/n$ behavior for
small densities, which is associated with the presence of the 
{}``Mexican hat'' dispersion \cite{next}, and for higher densities the
cyclotron mass increases as the carrier density increases. It is worth
mentioning that a more realistic calculation would account for the variations
of the bias with the density, $V(n)$, as varying one implies the variation of
the other. We postpone to future work the problem of the determination
of $V(n)$ \cite{next}. Also, the bias may not be the same for equal 
concentration of holes and electrons \cite{privateGeim, arpes}. In that case
the cyclotron effective mass is asymmetric with respect to doping with holes
or electrons. This is clearly seen in Fig.~\ref{cap:fig3}, where we show 
$m_\textrm{c}^+$ for $V/t = 0.05 \approx 135$~(meV)~--~holes~--~and 
$V/t = 0.005 \approx 13.5$~(meV)~--~electrons.

Landau level formation in bilayer graphene was already studied in the
continuum limit in Ref.~\cite{MF06}. Here we study the problem of a bilayer
subjected to a perpendicular magnetic field using the tight-binding
Hamiltonian given by Eq.~\eqref{eq:Hbilayer}, both at zero and finite bias.
Again we assume that inequalities $V \ll t_\perp \ll t$ hold.
Bilayer nano-ribbons with zigzag and armchair edges were diagonalized,
the obtained spectrum is shown in Fig.~\ref{cap:fig4}. For each $k$,
the momentum parallel to the edge, there are $4N$ bands. 
In Fig.~\ref{cap:fig4}, however, we focus on the low energy behavior near
the Dirac points. Panels~(a) and~(c) show the result for the unbiased bilayer
with zigzag and armchair edges, respectively. Fourfold degenerate zero
energy Landau levels are clearly seen, along with twofold degenerate
non-zero Landau levels, in agreement with the continuum result \cite{MF06}.
The major difference between zigzag and armchair is that the former have zero 
energy surface states along with zero energy bulk Landau levels \cite{next}. 
The result for the biased bilayer, $V = t_\perp /10=0.02$,
is shown in panels~(b) and~(d) for zigzag and armchair edges, respectively.
The fourfold degeneracy of zero energy Landau levels is lifted. A gap $V$
opens and twofold degenerate Landau levels at $V/2$ and $-V/2$ show up.
In fact it can be shown that their wave functions are either localized
in layer~1 or~2. For zigzag edges it becomes clear that surface states
and bulk Landau levels of the same band live in different layers \cite{next}.
As a consequence, dispersive surface states appear inside the gap,
as shown in Fig.~\ref{cap:fig4}~(d).

\section{Conclusions}
\label{sec:conclusions}
Using a tight-binding approximation we have studied the electronic behavior
of a graphene bilayer with layers at different electrostatic 
potential~--~{\it biased bilayer}. The applied bias opens a gap in the spectrum
which is completely controlled by the applied voltage. We have shown that
diagonal disorder reduces the size of the gap, which finally closes when the
width of the disorder distribution equals the larger energy scale in the
system: the in-plane hopping $t$. We have also studied the biased bilayer
in the presence of a perpendicular magnetic field. We have
calculated the semi-classical cyclotron effective mass as a function of the 
carrier density and bias, which is valid for low magnetic fields.
When the field is sufficiently high Landau level formation in zigzag and 
armchair nano-ribbons is perfectly seen, where a gap between
the lowest electron-like and the highest hole-like bulk Landau levels opens
in the presence of a finite bias.
\\
\\
We thank A.~H.~Castro~Neto and F.~Guinea for many illuminating discussions.
E.V.C. acknowledges the financial support of 
Funda\c{c}\~ao para a Ci\^encia e a Tecnologia 
through Grant  No.~SFRH/BD/13182/2003. J.M.B.L.S. and E.V.C. were additionally 
financed by FCT and EU through POCTI (QCAIII).
N.M.R.P. is thankful to the ESF Science Programme No.~INSTANS~2005-2010
and FCT and EU under the Grant No.~POCTI/FIS/58133/2004.

\end{document}